\documentclass[prb,superscriptaddress,showpacs]{revtex4}
\usepackage{graphicx}
\newcommand{\be}{\begin{equation}}
\newcommand{\ee}{\end{equation}}
\renewcommand\vec[1]{\mbox{\boldmath $#1$}}
\def\p{^{\,\prime}}

\newcounter{appendixc}
\renewcommand{\appendix}[1]{\vspace*{0.6cm}
\refstepcounter{appendixc} \setcounter{table}{0}
\setcounter{equation}{0}
\renewcommand{\thetable}{\Alph{appendixc}\arabic{table}}
\renewcommand{\theappendixc}{\Alph{appendixc}}
\renewcommand{\theequation}{\Alph{appendixc}\arabic{equation}}
\noindent{\bf Appendix \theappendixc. #1}\par\vspace*{0.4cm}}

\begin{document}

\title{Confined polar optical phonons in semiconductor double heterostructures: an improved continuum approach}
\author{F. Comas}
\thanks{On leave from: Departamento de F\'{\i}sica Te\'{o}rica, Universidad
de la Habana, Vedado 10400, Havana, Cuba}
\affiliation{Departamento de F\'{\i}sica, Universidade Federal de
S\~ao Carlos, 13565-905 S\~ao Carlos SP, Brazil}
\author{I. Camps}
\affiliation{Departamento de F\'{\i}sica, Universidade Federal de
S\~ao Carlos, 13565-905 S\~ao Carlos SP, Brazil}
\author{N. Studart}
\affiliation{Departamento de F\'{\i}sica, Universidade Federal de
S\~ao Carlos, 13565-905 S\~ao Carlos SP, Brazil}
\author{G. E. Marques}
\affiliation{Departamento de F\'{\i}sica, Universidade Federal de
S\~ao Carlos, 13565-905 S\~ao Carlos SP, Brazil}
\date{\today}

\begin{abstract}
Confined polar optical phonons are studied in a semiconductor
double heterostructure (SDH) by means of a generalization of a
theory developed some years ago and based on a continuous medium
model. The treatment considers the coupling of electro-mechanical
oscillations and involves dispersive phonons. This approach has
provided results beyond the usually applied dielectric continuum
models, where just the electric aspect of the oscillations is
analyzed. In the previous works on the subject the theory included
phonon dispersion within a quadratic (parabolic) approximation,
while presently linear contributions were added by a
straightforward extension of the fundamental equations. The
generalized version of the mentioned theoretical treatment leads
to a description of long wavelength polar optical phonons showing
a closer agreement with experimental data and with calculations
along atomistic models. This is particularly important for systems
where the linear contribution to dispersion becomes predominant.
We present a systematic derivation of the underlying equations,
their solutions for the bulk and SDH cases, providing us a
complete description of the dispersive modes and the associated
electron-phonon Hamiltonian. The results obtained are applied to
the case of a EuS/PbS/EuS quantum-well.
\end{abstract}

 \maketitle

\section{Introduction}

Polar optical phonons can be studied along the lines of continuous
media treatments for bulk semiconductors or semiconductor
microcrystallites leading to useful results valid in the long
wavelength limit \cite{c1,c2,c3}. This kind of approach has been
also applied in the case of semiconductor nanonostructures
(Quantum-Wells, Quantum-Wires, Quantum-Dots, etc.) and met an
acceptable degree of success in spite of the nanoscopic dimensions
of the systems \cite{c4,c5,c6,c7,c8,c9,c10}. The so-called
``dielectric continuous approach", based upon the well-known
Born-Huang equations, is frequently applied and  allows the
description of the LO, TO and IF (interface) polar optical
phonons. The main limitation of this approach should be related to
the pure electromagnetic nature of the treatment, while very often
just dispersionless phonons are addressed. Several years ago the
need for a better description of polar optical phonons was
recognized, while theories involving dispersive phonons and
discussing on equal grounds both the electric and mechanical
aspects of the oscillations were proposed. The matching boundary
conditions at the heterostructure interfaces were also
appropriately reanalyzed, putting forward one important issue: the
involved oscillations display a mixed character, and we cannot any
longer consider the TO, LO and IF phonons as individual entities.
Among the various proposed approaches, in
Refs.~[\onlinecite{c11,c12,c13}] a system of coupled differential
equations was derived involving the displacement vector $\vec{u}$
and the electric potential $\varphi$ for each segment of a given
heterostructure, while the boundary conditions at the interfaces
were also deduced from the fundamental equations in a cogent way.
The latter approach has been applied during more than a decade to
different nanostructures, meeting a reasonable degree of success
\cite{c14,c15,c16,c17,c18}. The obtained phonon modes are
dispersive and essentially valid near the Brillouin Zone (BZ)
center, while dispersion is assumed quadratic (parabolic) with the
wave vector $\vec{k}$.  However, it has been discussed that the
sole parabolic dispersion is not able to describe the phonons
within an acceptable approximation even if we limit ourselves to a
limited region of the BZ (near the $\Gamma$ point) \cite{c19}. The
latter issue is the main motivation for the present work.

We develop a generalization of the equations presented in the
above mentioned references allowing to introduce a phonon
dispersion that includes both linear and quadratic terms, thus
leading to a frequency dependence on $k$ of the general form
$\omega^2=\omega_0^2+\beta k+\beta\p k^2$ for the bulk
semiconductor. With this aim, the equations in the previous
references \cite{c11,c12,c13,c14} are modified. Concerning the
boundary conditions they need not to be altered, remaining the
same as in the cited references and in close correspondence with
the underlying equations. The new equations are solved for two
cases: the bulk semiconductor and the semiconductor double
heterostructure (SDH). The SDH is a well known system, in which a
certain semiconductor slab, made of material ``1'', is sandwiched
between two slabs of material ``2''. The results are applied to an
interesting case, the EuS/PbS/EuS QW, where it is evident that the
introduced generalizations play an important role It should be
emphasized that the added linear terms seem to be necessary for
achieving reliable agreement with experiment and atomistic
calculations in almost all the cases we have analyzed.

The paper is organized as follows. In Sec. II we propose the new
coupled differential equations involved in this approach and
discuss their physical meaning. Section III shows the solutions
for the bulk semiconductor case. In Section IV the solutions for
the SDH are derived, including a derivation of the three possible
oscillation modes and their dispersion laws. Section V is devoted
to normalization of the phonon states and a deduction of the
electron-phonon interaction Hamiltonian. Finally, in Section VI a
physical analysis of the EuS/PbS/EuS QW is done and some general
conclusions are also given.

\section{Fundamental Equations}

For the relative displacement vector $\vec{u}$ and the electric
potential $\varphi$  we assume the equations:

\be \label{1} \rho
(\omega^2-\omega_T^2)\vec{u}=\alpha\nabla\varphi+\rho\beta_L^2\nabla\nabla\cdot\vec{u}-\rho\beta_T^2\nabla\times\nabla\times\vec{u}+i\rho(\vec{a}\cdot\nabla)\vec{u}+i\rho\vec{b}(\nabla\cdot\vec{u})\,\,;
\ee and \be \label{2}
\nabla^2\varphi=\frac{4\pi\alpha}{\epsilon_{\infty}}\nabla\cdot\vec{u}
\;\;\;\; \mbox{with}\;\;\;\;
4\pi\alpha^2=\rho\omega_T^2(\epsilon_{0}-\epsilon_{\infty})=\rho\epsilon_{\infty}
(\omega_L^2-\omega_T^2)\;\;. \ee

In the Eqs.~(\ref{1}) and (\ref{2}) the same notation as in Ref.
\onlinecite{c11,c12,c13} is in general applied. $\rho$ is the
reduced mass density associated to the ion couple, $\epsilon_0$
($\epsilon_{\infty}$) is the low frequency (high frequency)
dielectric constant, $\omega_T$ ($\omega_L$)  is the transverse
(longitudinal) phonon frequency at the $\Gamma$ point of the bulk
semiconductor, and we assume the validity of the
Lyddane-Sachs-Teller relation
($\omega_L^2/\omega_T^2=\epsilon_0/\epsilon_{\infty}$). The
Maxwell equations are applied in the nonretarded limit ($c\to
\infty$) and the quantities are considered to depend harmonically
on the time ($\sim \exp (-i\omega t)$). Parameters $\beta_L$ and
$\beta_T$ (with dimensions of velocity) are introduced to describe
the quadratic (parabolic) phonon dispersion and used to fit the
bulk semiconductor dispersion curve. They were already considered
in previous works. Now we are introducing new parameters in the
form of vectors $\vec{a}$ and $\vec{b}$, not considered in
previous works and leading to linear terms in the phonon
dispersion law. Eqs.~(\ref{1}) and (\ref{2}) represent a system of
four coupled partial differential equations assumed to describe
the confined polar optical phonons in each segment of the
semiconductor heterostructure. The solutions must be matched at
the interfaces by applying the boundary conditions both for
$\vec{u}$ and $\varphi$. These boundary conditions were already
discussed in \cite{c11,c12,c13} and similar papers, and remain to
be valid in the present case as can be easily proved. Let us
recall them:

\begin{itemize}
\item Continuity of $\vec{u}$ and $\varphi$ at the interfaces.
\item Continuity of the normal component of the electric induction
vector $\vec{D}=\epsilon_{\infty}\vec{E}+4\pi\alpha\vec{u}$. \item
Continuity of the normal component of the ``stress tensor''
$\sigma_{ij}$, i.e., continuity of the force flux through the
interface.
\end{itemize}

In our case the tensor $\sigma_{ij}$ is defined in the form:
\[\sigma_{ij}=-\rho
(\beta_L^2-2\beta_T^2)(\nabla\cdot\vec{u})\delta_{ij}-\rho\beta_T^2\left(u_{j,i}+u_{i,j}\right)+i\rho\left
(b_iu_j+a_ju_i\right ) \;\;.\] where the suffixes
$i,\,j=1,\,2,\,3$ and should not be confused with the
$i=\sqrt{-1}$ used as a factor. Obviously, in all the equations
above we assume an isotropic model for the semiconductor compound.
The comma in $u_{i,\,j}$ denotes the partial derivative with
respect to $x_j$. The four latter terms at the r.h.s. of
Eq.~(\ref{1}) may be written as the divergence of tensor
$\sigma_{ij}$ and represent a certain force density which should
be added to the force density $-\rho\omega^2_T\vec{u}$. It must be
remarked that $\sigma_{ij}$, strictly speaking,  should not be
interpreted as the usual stress tensor applied in the theory of
elastic media. In the first place, vector $\vec{u}$ is the
relative displacement between the ions (and not the displacement
of each atom with respect to its equilibrium position). But, as
has been extensively discussed in all previously published works
on this subject, the considered terms are phenomenological in
their nature and  introduced with the direct purpose of
reproducing phonon dispersion as observed in experiments and in
microscopic (atomistic) calculations. In this connection tensor
$\sigma_{ij}$ must be considered as a pure phenomenological
quantity allowing a better description of phonon dispersion, and
we have not attempted to relate it to the elastic constants of a
continuous medium. However, it may be assumed to introduce in our
long wavelength approach an appropriate account of what is done in
atomistic treatments, where the interaction between and atom and
its different neighbors is used for the calculations.

We must emphasize that in some of the previously published papers
the third boundary condition for the tensor $\sigma_{ij}$ has not
been applied, because it was assumed that $\vec{u}|_S=0$ at the
interfaces. The latter boundary condition  holds whenever the
mechanical oscillations from a given segment of the
heterostructure do not show significant penetration into the
adjacent segments. This condition can be safely used when
semiconductor ``1'' does not vibrate mechanically  in the range of
mechanical vibrations of semiconductor ``2'', and this is well
fulfilled when the bandwidths of  the optical phonons belonging to
(bulk) semiconductors ``1'' and ``2'' is much smaller than the
frequency gap between them. Microscopic (atomic) models for the
optical phonons also support this approximation.  In this case the
second boundary condition reduces to the continuity of
$\epsilon_{\infty}\partial\varphi/\partial N|_S$ at each
interface.

\section{Bulk semiconductor case}

In the case of a bulk semiconductor the quantities $\vec{u}$ and
$\varphi$ should show a dependence on $\vec{r}$ in the form
$\sim\exp(i\vec{k}\cdot\vec{r})$, and direct substitution of these
solutions in Eqs.~(\ref{1}) and (\ref{2}) leads to:

\be \label{3} \varphi =-\frac{4\pi\alpha
i}{\epsilon_{\infty}k^2}(\vec{k}\cdot \vec{u}) \ee and \be
\label{4} \left(
\omega^2-\omega_T^2+\vec{a}\cdot\vec{k}+\beta_T^2k^2\right)\vec{u}=\left[
(\omega_L^2-\omega_T^2)\vec{k}/k^2-\beta_L^2\vec{k}+\beta_T^2\vec{k}-\vec{b}\right]
(\vec{k}\cdot\vec{u})\,\,. \ee

Taking $\vec{u}=\vec{u}_T+\vec{u}_L$ where
$\vec{k}\cdot\vec{u}_T=0$, $\vec{k}\cdot\vec{u}_L\neq 0$,
$\vec{k}\times\vec{u}_L=0$, and $\vec{k}\times\vec{u}_T\neq 0$ we
are directly led to:

\be \label{5}
(\omega^2-\omega_L^2+(\vec{a}+\vec{b})\cdot\vec{k}+\beta_L^2k^2)\vec{u}_L=0\;\;,
\ee and \be \label{6}
(\omega^2-\omega_T^2+\vec{a}\cdot\vec{k}+\beta_T^2k^2)\vec{u}_T=0\,\,.
\ee

Eq.~(\ref{5}) is obtained after scalar multiplication by vector
$\vec{k}$ and describes longitudinal mechanical vibrations coupled
to the electric potential $\varphi$. Eq.~(\ref{6}) follows after
vector multiplication by $\vec{k}$ and describes  transversal
mechanical vibrations not involving an electric field. The
corresponding dispersion laws are evidently implicit in the
equations. Moreover, by applying the equation
$\vec{D}=\epsilon(\omega,\,\vec{k})\vec{E}=\epsilon_{\infty}\vec{E}+4\pi\alpha\vec{u}$,
we can find the dielectric function in the form:

\be \label{7} \frac{\epsilon(\omega,
\,\vec{k})}{\epsilon_{\infty}}=\frac{\omega^2-\omega_L^2+(\vec{a}+\vec{b})\cdot\vec{k}+\beta_L^2k^2}{\omega^2-\omega_T^2+\vec{a}\cdot\vec{k}+\beta_T^2k^2}\,,
\ee which is the expected structure of the dielectric function for
this type of dispersive phonons.

Let us now remark that, within the isotropic model we are
applying, the direction of vectors $\vec{a}$ and $\vec{b}$ may be
chosen in a suitable way. Thus, we have found convenient to take
them along the vector $\vec{k}$, so we can finally write
$\vec{a}\cdot\vec{k}=ak$ and $\vec{b}\cdot\vec{k}=bk$. The sign
and value of the quantities $a$ and $b$ should be found after a
fitting of the bulk semiconductor dispersion law, embracing a
certain percent (say, 30-40 \%) of the BZ near the $\Gamma$ point.
By the same method parameters $\beta_{T(L)}$ should be determined.
Depending on the involved semiconductor compound we may be led to
consider the quantity $\beta_{T(L)}^2$ with a minus sign, and this
can be easily done by the formal substitution $\beta_{T(L)}\to
i\beta_{T(L)}$.

\section{Semiconductor Double Heterostructure case}

We consider the SDH grown along the z-axis, such that for
$-d/2<z<d/2$ we have semiconductor ``1'', and for $|z|>d/2$ we
have semiconductor ``2''. Obviously, the interfaces are located at
$z=\pm d/2$ and are formally assumed as infinite plane surfaces.
In order to find the solutions of Eqs.~(\ref{1}) and (\ref{2}) we
shall use the auxiliary potentials $\vec{A}$ and $\Psi$, such that:

\be \label{8} \vec{u}=\nabla \Psi + \nabla\cdot\vec{A} \quad
\mbox{with} \quad \nabla\cdot\vec{A}=0\,\,. \ee

Substitution of Eq.~(\ref{8}) in Eqs.~(\ref{1}) and (\ref{2}),
provides equations for $\Psi$ and $\vec{A}$ having the form:

\be \label{9} \nabla^2 \left
[\nabla^2\Psi+\frac{\omega_L^2-\omega^2}{\beta_L^2}\Psi+\frac{i}{\beta_L^2}(\vec{a}+\vec{b})\cdot\nabla\Psi
\right ]=0\;\;, \ee and \be \label{10} \nabla^2 \left
[\nabla^2\vec{A}+\frac{\omega_T^2-\omega^2}{\beta_T^2}\vec{A}+\frac{i}{\beta_T^2}\vec{a}\cdot\nabla\vec{A}+\frac{i}{\beta_T^2}\vec{b}\times\nabla\Psi
\right ]=0\;\;, \ee As may be seen the Eqs.~(\ref{9}) and
(\ref{10}) bear a certain analogy with the corresponding equations
in Ref. \onlinecite{c14}. However, besides the presence of linear
terms in $\nabla$ involving first order derivatives, we notice
that Eq.~(\ref{10}) is now coupled to Eq.~(\ref{9}). We are thus
led to equations showing some significant differences with respect
to those of  \cite{c14}.

In order to find the solutions of the Eqs.~(\ref{9}) and
(\ref{10}) we consider the wave vector $\vec{\kappa}\equiv
(\kappa_x,\,\kappa_y)$, belonging to the $(x,\,y)$ plane, and
propose:

\be \label{11} \Psi (\vec{r})=v(z)\exp (i\vec{\kappa}\cdot
\vec{R}) \quad , \quad \vec{A}(\vec{r})=\vec{A}(z)\exp
(i\vec{\kappa}\cdot \vec{R})\,\,. \ee

The exponential in Eq.~(\ref{11}) is a consequence of the
translational invariance within the $(x,\,y)$ plane, this
invariance is evidently destroyed for displacements along the $z$
axis. We are taking $\vec{R}\equiv (x,\,y)$. Vectors $\vec{a}$ and
$\vec{b}$ shall be taken along vector $\vec{\kappa}$. Thus, we are
led to the following equation for $v(z)$:

\be \label{12} \frac{d^2v}{dz^2}+Q^2_Lv=s \quad \mbox{with}\quad
Q^2_L=\frac{\omega_L^2-\omega^2-(a+b)\kappa}{\beta_L^2}-\kappa^2\,\,,
\ee where $s$ satisfies the equation $\left[
\frac{d^2}{dz^2}-\kappa^2\right ]s=0$. It is then obvious that the
general solution for $v(z)$ is given in the form:

\be \label{13} v(z)=B\exp (iQ_Lz)+B\p\exp
(-iQ_Lz)+\frac{1}{Q^2_L+\kappa^2}\left(C\exp (\kappa z)+C\p\exp
(-\kappa z)\right )\,\,, \ee displaying four constants as
corresponds to a solution of Eq.~(\ref{9}). The latter solution
applies to the interval $-d/2<z<d/2$, while for $|z|>d/2$ the
divergent solutions should be eliminated.

In our model of an isotropic semiconductor, we may take advantage
of rotational invariance about the $z$ axis, and, without lost of
generality, take the $y$ axis along vector $\vec{\kappa}$. Hence,
we can write $\vec{\kappa}\equiv (0,\,\kappa)$. Under the latter
conditions, the equation for $\vec{A}(z)$ is reduced to:

\be \label{14}
\frac{d^2\vec{A}}{dz^2}+Q_T^2\vec{A}(z)+\frac{ib}{\beta_T^2}\frac{dv}{dz}\vec{e}_z=\vec{s}\quad
\mbox{with}\quad
Q_T^2=\frac{\omega_T^2-\omega^2-a\kappa}{\beta_T^2}-\kappa^2\,\,,
\ee where the components of vector $\vec{s}$ are solutions of the
equation $\left [\frac{d^2}{dz^2}-\kappa^2\right ]s_i=0$ and $s_i$
with $i=1,\,2,\,3$ represents the cartesian components. Vector
$\vec{e}_z$ is a unit vector along the $z$ axis. We are thus led
to the following equations for the components of $\vec{A}$:

\be \label{15} \left[ \frac{d^2}{dz^2}+Q_T^2\right ]
A_x+\frac{ib}{\beta_T^2}\frac{dv}{dz} =s_x \quad , \quad \left[
\frac{d^2}{dz^2}+Q_T^2\right ] A_y =s_y \quad , \quad \left[
\frac{d^2}{dz^2}+Q_T^2\right ] A_z =s_z\;\;. \ee

The solutions of Eqs.~(\ref{15}) are given in the form:

\[A_x=B_x\exp (iQ_Tz)+B_x\p\exp (-iQ_Tz)+\frac{bQ_L}{\beta_T^2(Q_T^2-Q_L^2)}\left [B\exp(iQ_Lz)-B\p\exp(-iQ_Lz)\right ]\]
\be \label{18} +\frac{1}{Q_T^2+\kappa^2}\left [(C_x-i\delta
C\p)\exp (\kappa z) +(C_x\p+i\delta C\p)\exp (-\kappa z)\right
]\;\;, \ee where $\delta=b\kappa/(\beta_T^2(Q_L^2+\kappa^2))$. \be
\label{19} A_j=B_j\exp (iQ_Tz)+B_j\p\exp
(-iQ_Tz)+\frac{1}{Q_T^2+\kappa^2)}\left [C_j\exp(\kappa
z)+C_j\p\exp(-i\kappa z)\right ] \quad \mbox{with} \quad j=y,\;z
\ee

The condition $\nabla\cdot\vec{A}=0$ leads to the equation
$i\kappa A_y(z)+\frac{d}{dz}A_z(z)=0$ and gives additional
relations for the constants: $\kappa B_y=-Q_TB_z$, $\kappa
B_y\p=Q_TB_z\p$, $iC_y=-C_z$, and $iC_y\p=C_z\p$.

Using the latter results for $\Psi$ and $\vec{A}$ in
Eq.~(\ref{8}), we may obtain an explicit expression for $\vec{u}$.
We are thus led to:

\be \label{20} \vec{u}=\left
[X(z)\vec{e}_x+Y(z)\vec{e}_y+Z(z)\vec{e}_z\right ]\exp(i\kappa y)
\;\;, \ee where
\begin{eqnarray}
X(z)&=&iQ_T\left [B_y\exp(iQ_Tz)-B_y\p\exp(-iQ_Tz)\right ]\;\;,\label{21}\\
Y(z)&=&-iQ_T\left [B_x\exp(iQ_Tz)-B_x\p\exp(-iQ_Tz)\right ]+ik\Gamma \left[ B\exp(iQ_Lz)+B\p\exp(-iQ_Lz)\right]\nonumber\\
       &+  &\frac{i\kappa}{Q_T^2+\kappa^2}\left[C\exp(\kappa z)+C\p\exp(-\kappa z)\right ]\label{22}\;\;,\\
Z(z)&=&i\kappa \left [B_x\exp(iQ_T z)+B_x\p\exp(-iQ_Tz)\right ]+iQ_L\Gamma\p\left [B\exp(iQ_Lz)-B\p\exp(-iQ_Lz)\right ]\nonumber\\
&+  &\frac{\kappa}{Q_T^2+\kappa^2}\left [C\exp(\kappa
z)-C\p\exp(-\kappa z)\right ] \label{23}
\end{eqnarray}
with
\[ \Gamma=1+\frac{bQ_L^2}{\beta_T^2\kappa (Q_T^2-Q_L^2)}\quad ,\quad \Gamma\p=1-\frac{b\kappa}{\beta_T^2 (Q_T^2-Q_L^2)} \]
while the constants were redefined in a convenient way. Once we
have the solutions for $\vec{u}$, the solution for the electric
potential is easily found by using Eq.~(\ref{2}). We just write
the final result:

\be \label{24}
\varphi=\left\{\frac{4\pi\alpha}{\epsilon_{\infty}}\left
[B\exp(iQ_Lz)+B\p\exp(-iQ_Lz)\right]-C\exp(\kappa
z)-C\p\exp(-\kappa z)\right \}\exp(i\kappa y)=f(z)\exp(i\kappa
y)\;\;. \ee

It may be seen that the $X$ solutions represent uncoupled
oscillations of a pure transversal character and not involving an
electric potential. However, solutions for $Y$, $Z$ and $\varphi$
are coupled solutions. Up to this point we have not taken into
account the system invariance under the transformation $z\to -z$.
If the latter symmetry is used, the solutions can be classified
according to their parity. Then, for the coupled solutions, we can
write:

{\bf Case (a)}:   \underline{Odd Potential States (OPS)}

\begin{eqnarray}
\frac{Y(z)}{C}&=&-\kappa \Gamma S_1\sin (Q_Lz)+Q_TS_2\sin(Q_Tz)+S_3\sinh(\kappa z)\;\;,\nonumber\\
\frac{Z(z)}{C}&=&iQ_L\Gamma\p S_1\cos(Q_Lz)+i\kappa S_2\cos(Q_Tz)-iS_3\cosh(\kappa z)\;\;,\nonumber\\
\frac{f(z)}{C}&=&\frac{4\pi\alpha}{\epsilon_{1\infty}}i\left
[S_1\sin(Q_Lz)+\frac{\beta_T^2(Q_L^2+\kappa^2)}{\kappa
(\omega_L^2-\omega_T^2)}S_3\sinh(\kappa z)\right]\;\;,\label{25}
\end{eqnarray}

In the Eqs.~(\ref{25}) we have required $\vec{u}=0$ at the
interfaces. The solutions given are defined for $-d/2<z<d/2$. For
$|z|>d/2$ we considered $\vec{u}\equiv 0$ while the solution for
$f(z)$ is casted as:
\[f(z)=\frac{4\pi\alpha i}{\epsilon_{1\infty}}C\exp(\nu )\left [S_1\sin\eta+\frac{\beta_T^2(Q_T^2+\kappa^2)}{\kappa (\omega_L^2-\omega_T^2)}S_3\sinh \nu \right ]\frac{z}{|z|}\exp(-\kappa |z|)\;\;,\]
and the continuity of the potential at the interfaces is ensured.
We are using the notation: $\nu = \kappa d/2$, $\eta =Q_Ld/2$ and
also $\mu = Q_Td/2$. The functions $S_i$ ($i=1,\;2,\;3$) are
reported in the Appendix. The dispersion relation for this type of
oscillations is obtained from the boundary condition
$\epsilon_{1\infty}\partial f_</\partial
z|_{d/2}=\epsilon_{2\infty}\partial f_>/\partial z|_{d/2}$ (the
notation $f_{<(>)}$ is useful to denote the function at each side
of the interface) and may be written as follows:

\[\gamma \left [(\epsilon_{1\infty}/\epsilon_{2\infty})\eta\cos\eta+\nu\sin\eta \right]\left[\nu\sinh\nu\cos\mu+\mu\cosh\nu\sin\mu\right]=\]
\begin{equation} \label{26}
-(\mu^2+\nu^2)\left[(\epsilon_{1\infty}/\epsilon_{2\infty})\cosh\nu+\sinh\nu\right]\left[\Gamma\nu^2\sin\eta\cos\mu+\Gamma\p\mu\eta\sin\mu\cos\eta\right]\;\;,
\end{equation}
where the parameter $\gamma$ has the form:
\[\gamma=\frac{(\omega_L^2-\omega_T^2)d^2}{4\beta_T^2}\] .

In the limit $\kappa \to 0$ Eq.~(\ref{26}) reduces to:
$\sin\mu\cos\eta =0$ and the zeros are simply given by: $\mu =
N\pi$ and $\eta = (2N+1)\pi/2$.

{\bf Case (b)}:   \underline{Even Potential States (EPS)}

In this case we have

\begin{eqnarray}
\frac{Y(z)}{C\p}&=&i\kappa \Gamma S_1\p\cos (Q_Lz)-iQ_TS_2\p\cos(Q_Tz)+iS_3\p\cosh(\kappa z)\;\;,\nonumber\\
\frac{Z(z)}{C\p}&=&-Q_L\Gamma\p S_1\p\sin(Q_Lz)-\kappa S_2\p\sin(Q_Tz)+S_3\p\sinh(\kappa z)\;\;,\nonumber\\
\frac{f(z)}{C}&=&\frac{4\pi\alpha}{\epsilon_{1\infty}}\left
[S_1\p\cos(Q_Lz)-\frac{\beta_T^2(Q_L^2+\kappa^2)}{\kappa
(\omega_L^2-\omega_T^2)}S_3\p\cosh(\kappa z)\right]\;\;,\label{27}
\end{eqnarray}
valid for $-d/2<z<d/2$. For $|z|>d/2$ $\vec{u}\equiv 0$ and
\[f(z)=\frac{4\pi\alpha }{\epsilon_{1\infty}}C\exp(\nu )\left [S_1\p\cos\eta-\frac{\beta_T^2(Q_T^2+\kappa^2)}{\kappa (\omega_L^2-\omega_T^2)}S_3\p\cosh \nu \right ]\exp(-\kappa |z|)\;\;.\]

The functions $S_i\p$ ($i=1,\,2,\,3$) are given in the Appendix.
The dispersion law, obtained in the same way than in the former
case, reads:

\[\gamma\left[(\epsilon_{1\infty}/\epsilon_{2\infty})\eta\sin\eta-\nu\cos\eta\right]\left[\mu\sinh\nu\cos\mu-\nu\cosh\nu\sin\mu\right]=\]
\begin{equation}\label{28}
-(\mu^2+\nu^2)\left[(\epsilon_{1\infty}/\epsilon_{2\infty})\sinh\nu+\cosh\nu\right]\left[\Gamma\nu^2\cos\eta\sin\mu+\Gamma\p\mu\eta\\cos\mu\sin\eta\right]
\end{equation}

In the limit $\kappa \to 0$ Eq.~(\ref{28}) reduces to:
$\cos\mu\sin\eta =0$ and the zeros are simply given by: $\eta =
N\pi$ and $\mu = (2N+1)\pi/2$. Let us also remark that, both for
the OPS and the EPS, the following relation holds: \be\label{28a}
\mu^2=\left (\frac{\beta_L}{\beta_T}\right )^2\eta^2+\left
(\frac{bd}{2\beta_T^2}\right )\nu+\left
(\frac{\beta_L^2}{\beta_T^2}-1\right )\nu^2-\gamma\;\;, \ee
indicating that $\mu$ and $\eta$ are not independent quantities.

\begin{figure*}[tbp]
\includegraphics*{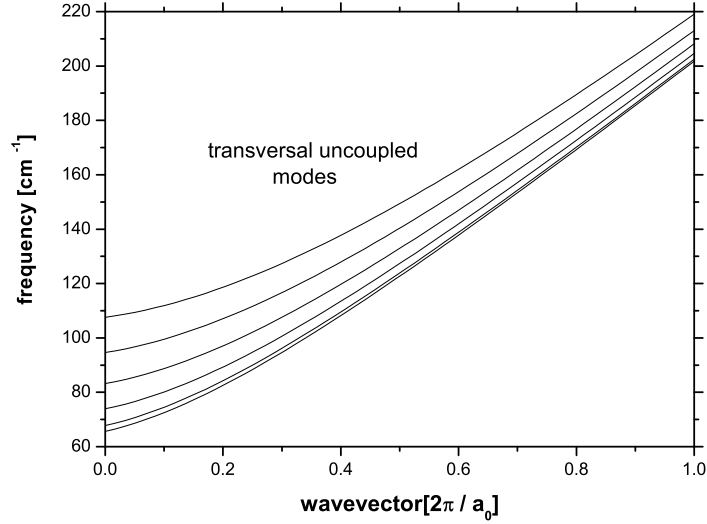}
\caption{Frequency (in units of cm$^{-1}$) as a function of the
wave vector $\kappa$ (in units of $2\pi /a_0$). We display the
first six modes. \label{1}}
\end{figure*}

{\bf Case (c)}: \underline{Uncoupled Modes}

As remarked above, these modes are related to the $x$ component of
$\vec{u}$, and should be classified as follows. The even uncoupled
modes (EUM) and the odd uncoupled modes (OUM), described by:

\be \label{28b} X(z)=\tilde{C}\cos(Q_Tz)\quad \mbox{(EUM)}\quad ;
\quad X(z)=\tilde{C}\p\sin(Q_Tz) \quad \mbox{(OUM)}\;\;, \ee where
the corresponding dispersion laws are:

\begin{eqnarray}
\mbox{(EUM)}\;\;\cos\mu=0 \quad \mbox{with}\quad \omega^2=\omega_T^2-a\kappa-\beta_T^2\kappa^2-\left (\frac{N_{odd}\pi\beta_T}{d}\right )^2 \;\;,\nonumber\\
\mbox{(OUM)}\;\;\sin\mu=0 \quad \mbox{with}\quad
\omega^2=\omega_T^2-a\kappa-\beta_T^2\kappa^2-\left
(\frac{N_{even}\pi\beta_T}{d}\right )^2 \label{28bb} \;\;,
\end{eqnarray}
where $N_{even}$ ($N_{odd}$ ) denotes an even (odd) integer.

\section{Normalization and Electron-Phonon Hamiltonian}

For the normalization of the oscillations we shall transform the
classical field $\vec{u}$ into a quantum-field operator by means
of the substitution $\vec{u}\to
\hat{\vec{u}}=\vec{u}\hat{a}_{\kappa\,,\,n}$, where
$\hat{a}_{\kappa\,,\,n}$ are second quantization operators obeying
boson commutation rules. The classical kinetic energy for the
oscillations should also be transformed into a quantum-mechanical
operator as follows:

\be \label{29} W_{kin}=\frac{1}{2}\rho\int_V
\dot{\vec{u}}^{*}\cdot\dot{\vec{u}}\,d^3r=\frac{1}{2}\rho\omega^2\int_V
\vec{u}^{*}\cdot{\vec{u}}\,d^3r\;\to \hat{H}_{ph}\;, \ee where \be
\label{30} \hat{H}_{ph}=\frac{1}{4}\rho\omega^2\int_V \left
(\vec{u}^{\dagger}\cdot\vec{u}+\vec{u}\cdot{\vec{u}}^{\dagger}\right
)\,d^3r=\frac{1}{2}\rho\omega^2\int_V\vec{u}^{*}\cdot\vec{u}\,d^3r\,\left
(\hat{a}_{\kappa\,,\,n}^{\dagger}\hat{a}_{\kappa\,,\,n}+\frac{1}{2}\right
)\;. \ee In Eq.~(\ref{30}) $\hat{H}_{ph}$ describes the free
phonons hamiltonian operator, while its hermitian character is
ensured by construction. Requiring the latter expression to be
given in standard form for phonons of the type $\kappa \,,\,n$
(actually, we must have $\omega \to \omega_{\kappa\,,\,n}$)
$\hat{H}_{ph}=\hbar\omega_{\kappa\,,\,n}\left
(\hat{a}_{\kappa\,,\,n}^{\dagger}\hat{a}_{\kappa\,,\,n}+\frac{1}{2}\right
)$, one can finally find the normalization constant $C$:

\be \label{31}
C_{\kappa\,,\,n}=\sqrt{\frac{2\hbar}{S\rho\omega_{\kappa\,,\,n}M_{\kappa\,,\,n}}}\;,
\ee where $S$ is a normalization area (within the $(x,\,y)$ plane)
and $M_{\kappa\,,\,n}=\int (Y^{*}(z)Y(z)+Z^{*}(z)Z(z))dz/|C|^2$.
The explicit expressions for $M_{\kappa\,,\,n}$ can be
analytically derived and are different for the OPS or the EPS.
They are given in the Appendix.

Let us now deduce the electron-phonon interaction hamiltonian. We
need to transform the electric potential into a quantum-mechanical
operator by means of the substitution: $\varphi \to
\hat{\varphi}=\varphi\hat{a}_{\kappa\,,\,n}$. According to
Eqs.~(\ref{25}) or (\ref{27}) the electric potential displays the
general form \[\varphi
=\frac{4\pi\alpha}{\epsilon_{1\infty}}CF(z)\exp(i\vec{\kappa}\cdot\vec{R})\;\;,\]
where the explicit expressions for $C$ and $F(z)$ depend on the
type of phonons (OPS or EPS). Now, the electron-phonon interaction
hamiltonian may be derived by means of
$\hat{H}_{e-ph}=-e(\varphi\hat{a}_{\kappa\,,\,n}+\varphi^{*}\hat{a}_{\kappa\,,\,n}^{\dagger})/2$.
Considering the Eq.~(\ref{31}) and avoiding immaterial phase
factors we are led to:

\be \label{32}
\hat{H}_{e-ph}=\sum_{\kappa\,,\,n}\sqrt{\frac{2\pi\hbar
e^2(\omega_L^2-\omega_T^2)}{S\epsilon_{1\infty}\omega_{\kappa\,,\,n}M_{\kappa\,,\,n}}}\left
[
F(z)_{\kappa\,,\,n}\exp(i\vec{\kappa}\cdot\vec{R})\,\hat{a}_{\kappa\,,\,n}+h.c.\right]\;,
\ee where ``h.c.'' stands for ``hermitian conjugate'' and the
quantities $M_{\kappa\,,\,n}$ and $F_{\kappa\,,\,n}(z)$ are
different for the OPS and the EPS  (and reported in the Appendix).
The latter electron-phonon hamiltonian is obviously a function of
$\vec{R}$ and $z$, and involves a summation over all possible
phonon  modes.

\begin{figure*}[tbp]
\includegraphics*{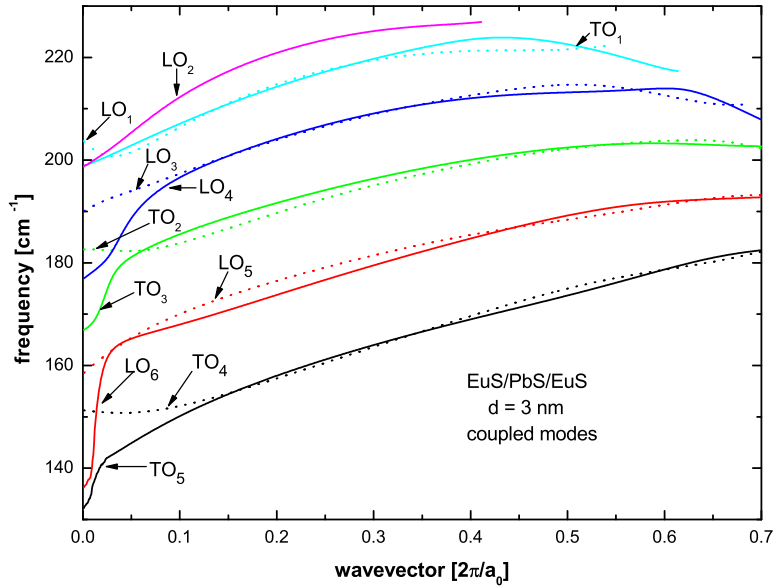}
\caption{Frequency (in units of cm$^{-1}$) as a function of the
wave vector $\kappa$ (in units of $2\pi /a_0$). We display eleven
coupled phonon modes, corresponding to the OPS (dotted curves) and
the EPS (continuous curves) oscillations. The curves were
identified by their behavior at $\kappa =0$. These confined
phonons represent oscillations of the PbS layer for $d=3$ nm.
\label{2}}
\end{figure*}

\section{Discussion of the results obtained}

In order to illustrate the physical meaning of the foregoing
results, we shall discuss a EuS/PbS/EuS SDH, a system which has
received attention in the latter years. The EuS is a large-gap
($E_{gap}=1.6$ eV) magnetic semiconductor behaving as a
ferromagnetic compound below $T_C=16.5$ K, while PbS is a narrow gap
($E_{gap}=0.3$ eV) semiconductor. Both materials grow in the rock
salt structure and show a rather good lattice matching ($\Delta
a/a \sim 0.6\,\%$). From the viewpoint of the electronic states
the layers of PbS behave as quantum-wells and the EuS layers
represent barriers \cite{c21}. These types of heterostructures are
complicated for the calculation of electronic states due to non
parabolicity and anisotropy of the band structure \cite{c21,c22}.
In the current work we provide the confined polar optical modes of
PbS for the mentioned heterostructure as a direct application of
our theoretical results. Concerning the phonon states the europium
chalcogenides have not been very well studied because of very high
neutron absorption. However, reliable theoretical calculations for
the bulk semiconductor can be found in Ref. \onlinecite{c23}. From
this reference we have taken the parameters
$\epsilon_{\infty}=4.9$, $\omega_{T}(\Gamma)=178.4\;\;cm^{-1}$ and
$\omega_{L}(\Gamma)=267.0\;\;cm^{-1}$ valid for EuS. For the PbS
the corresponding parameters are:
$\omega_T(\Gamma)=65.6\,\,cm^{-1}$,
$\omega_L(\Gamma)=205.6\,\,cm^{-1}$, $a_0=5.92\times 10^{-8}\,cm$
(the PbS lattice parameter), $\epsilon_{1\infty}=18.4$,
$a=-2.198\times10^{18}$ cm/$s^2$, $b=-1.502\times10^{19}$
cm/$s^2$, $\beta_L=471934$ cm/s, $\beta_T=306568\sqrt{-1}$ cm/s.
The values of $\beta_{L(T)}$, $a$, and $b$ were determined by the
authors after fitting the bulk PbS optical phonon dispersion laws
\cite{c24} covering approximately 30-40  $\%$  of the BZ near the
$\Gamma$ point. The values of $\omega_{T(L)}(\Gamma)$ were also
provided by the same reference, while $\epsilon_{\infty}$ comes
from Ref. \onlinecite{c25} (see also Ref. \onlinecite{c26}). We
can see a large difference between the $\omega_T(\Gamma)$ values
for the two semiconductors. The corresponding difference for the
$\omega_L(\Gamma)$ frequencies is not that large, but in both
cases (see, for instance, Fig. 4 of Ref. \onlinecite{c23}) the
band widths of bulk L or T phonons are relatively narrow. Under
these circumstances, as discussed in the Section 2, the validity
of  the boundary conditions we have used ($\vec{u}|_S=0$) is well
satisfied.

Let us first consider the so-called uncoupled modes, described by
Eqs.~(\ref{28b}) and (\ref{28bb}). In the Fig.~1 we can see the
frequencies of the first six uncoupled modes as a function of the
wave vector. These modes represent pure mechanical oscillations of
transversal type and do not involve electric potentials. Their
interaction with the electrons is weak and is determined through
the deformation potential. Their physical importance should be
related to the actual possibility of detecting these oscillations
by means of optical experiments, specially by IR spectroscopy.
Such information is of use in the investigation of the
nanostructure and may be related to geometrical characteristics of
the system.
\begin{figure*}[tbp]
\includegraphics*{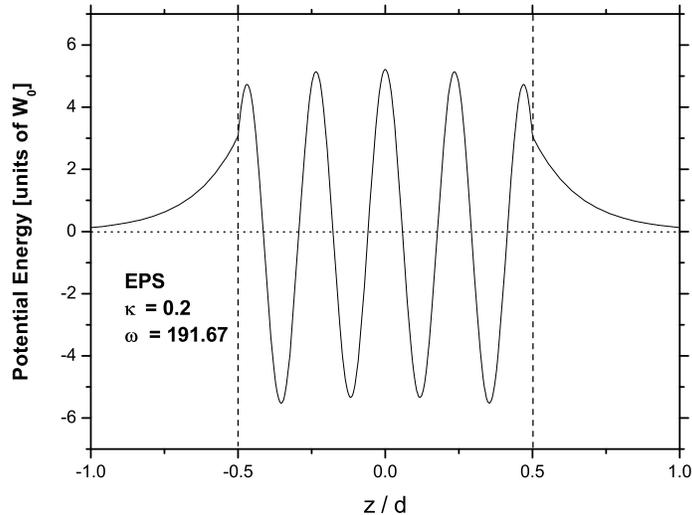}
\caption{Electric potential energy (in units of $W_0$) as a
function of $z$. We have chosen $\kappa = 0.2$ (units of
$2\pi/a_0$) and $\omega = 191.67$ cm$^{-1}$, corresponding to an
EPS.  \label{3}}
\end{figure*}

The so-called ``coupled modes'' are more interesting, involving a
richer physical structure and are Raman active. These modes, as
explained in the text above, are coupled to a long-range electric
potential and were classified as OPS and EPS. In Fig.~2 we are
showing the frequencies of the coupled modes (in cm$^{-1}$) as a
function of the wave vector $k$ (in units of $2\pi/a_0$). We have
considered a EuS/PbS/EuS SDH with $d=3$ nm, and the phonons
displayed in the figure correspond to oscillations within the PbS
layer, i.e., they are polar optical phonons confined to this
layer. The curves were obtained by numerical solution of
Eq.~(\ref{26}) for the OPS and Eq.~(\ref{28}) for the EPS. In
Fig.~2 eleven modes are displayed and denoted by LO$_i$ or TO$_i$
according to their behavior at $\kappa =0$. The OPS modes are
shown by dotted curves, while the EPS modes correspond to the
continuous curves. Of course, curves corresponding to modes of
different parity can cross each other. As we may see in the figure
the dotted curves freely cross the continuous ones and, for some
modes, practically show a high degree of overlapping. However,
curves of the same parity do not cross each other and, when they
become too close, anticrossing effects are seen. This is the case
of the modes LO$_6$ and TO$_5$, which create an anticrossing
effect near $\kappa =0$. In Fig.~2 we present the modes for a
frequency interval from 130 up to 230 cm$^{-1}$, where the
mentioned eleven modes were found. Actually, the number of
possible phonon modes is limited by the thickness of the PbS
layer, and we are probably showing more modes than those really
possible. An important characteristic of the confined coupled
phonons is  the mixed nature they show. We no longer have pure TO,
LO or IF phonons, but the dispersion curves may be predominantly
TO, LO or IF for different values of the wave vector.

\begin{figure*}[tbp]
\includegraphics*{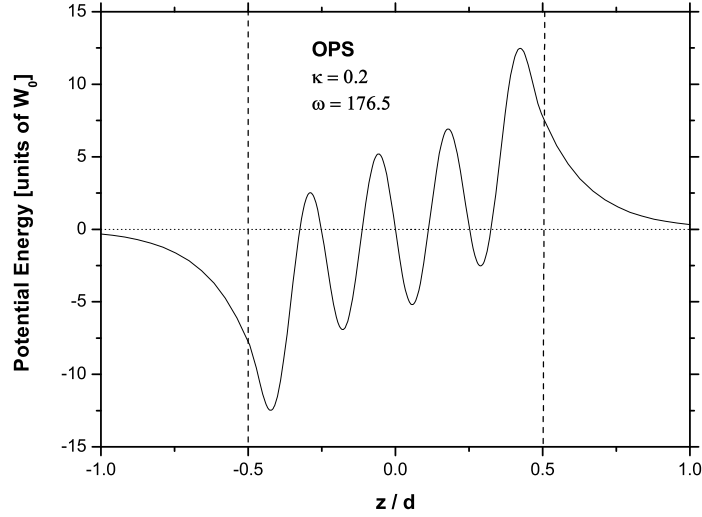}
\caption{Electric potential energy (in units of $W_0$) as a
function of $z$. We have chosen $\kappa = 0.2$ (units of
$2\pi/a_0$) and $\omega = 176.5$ cm$^{-1}$, corresponding to an
OPS. \label{4}}
\end{figure*}

However, at $\kappa = 0$ they recover their pure TO or LO nature,
a fact that was helpful in other to label the curves. The regions
showing a strong change in the curve slope (mainly near $\kappa
=0$) represent modes where the interface character is more
significant, involving larger electric potentials. It is
interesting to remark that some curves are interrupted (see the
modes LO$_1$, LO$_2$, and TO$_1$) for large values of $\kappa$.
This effect is found in the numerical computations, and may be
interpreted in terms of the applied approximation. The
approximation we are using in this work has a limited range of
validity and should provide reliable results for wave vectors just
near the BZ $\Gamma$ point. In Fig.~2 we are extrapolating our
results well beyond this region and, therefore, we should expect
our equations to fail for large wave vectors, especially if the
frequencies are high.

In Figs.~3 and 4 we displayed the electric potential energy
associated to the vibrations for fixed values of $\kappa$ and
$\omega$. We have selected two different parities: EPS for Fig.~3
and OPS for Fig.~4. These quantities were plotted in units of
$W_0=\left [2\pi \hbar e^2
(\omega_L^2-\omega_T^2)d/\epsilon_{1\infty}\omega_T S\right
]^{1/2}$ as a function of $z$ in the interval $-d<z<d$. They show
an oscillatory behavior within the QW region and decay
exponentially for values of $z$ in the outside region. The
potential energy as a function of $z$, and avoiding its $y$ and
$x$ dependence, gives us a qualitative estimation of the
electron-phonon interaction in the SDH. However, it should be kept
in mind that the electron wave function also plays a significant
role in the calculation of the interaction. When we study
quantities like scattering rates or polaronic effects the
interaction matrix elements must be calculated. In these cases the
oscillatory character of the interaction hamiltonian could be
responsible for a weaker interaction, but a final conclusion is
possible just after effectively making the calculations.

In conclusion, the main contribution of the present work is to
provide a long wavelength treatment of confined polar optical
phonons for SDH, which has been improved with respect to previous
treatments described in
Refs.~\onlinecite{c11,c12,c13,c14,c15,c16,c17,c18} in the sense it
involves a better description of phonon dispersion near the BZ
$\Gamma$ point. The inclusion of linear terms together with the
already considered quadratic ones appear  to be of importance for
almost all semiconductor compounds, leading to results of higher
reliability. In the especial case of the EuS/PbS/EuS SDH this
improvement plays an important role, and may be related to the
results reported in Ref.~\onlinecite{c19}.

\begin{acknowledgments} The work is partially supported by grants from the
Funda\c{c}\~ao de Amparo \`a Pesquisa de S\~ao Paulo and Conselho
Nacional de Desenvolvimento Cient\'{i}fico e Tecnol\'{o}gico. F.C.
is grateful to Departamento de F\'{\i}sica, Universidade Federal
de S\~ao Carlos, for hospitality.
\end{acknowledgments}
\appendix{Some of the used functions}

In the OPS (Eqs.~(\ref{25})) we used the following functions:

\begin{eqnarray}
S_1&=&\kappa \cos\mu\sinh\nu+Q_T\sin\mu\cosh\nu\;\;,\\
S_2&=&\kappa\Gamma\sin\eta\cosh\nu-Q_L\Gamma\p\cos\eta\sinh\nu\;\;,\\
S_3&=&\kappa^2\Gamma\cos\mu\sin\eta+Q_TQ_L\Gamma\p\sin\mu\cos\eta\;\;.
\end{eqnarray}

For the EPS (Eqs.~(\ref{28})) the corresponding functions are:

\begin{eqnarray}
S_1\p&=&Q_T\cos\mu\sinh\nu-\kappa \sin\mu\cosh\nu\;\;,\\
S_2\p&=&\kappa\Gamma\cos\eta\sinh\nu+Q_L\Gamma\p\sin\eta\cosh\nu\;\;,\\
S_3\p&=&\kappa^2\Gamma\sin\mu\cos\eta+Q_TQ_L\Gamma\p\cos\mu\sin\eta\;\;.
\end{eqnarray}

The quantity $M_{\kappa\,,\,n}$ introduced in  Eq.~(\ref{31})
involves tediously long but straightforward calculations, and
should be calculated independently for the OPS and the EPS. Here
we just report final results. For the OPS we obtained:

\begin{eqnarray}
M^{(OPS)}_{\kappa\,,\,n}&=&\frac{2}{d}S_1^2(\Gamma^2\nu^2+\Gamma^{'2}\eta^2)+\frac{2}{d}S_2^2(\mu^2+\nu^2)+\frac{2}{d}S_1^2(\Gamma^{'2}\eta^2-\Gamma^2\nu^2)\frac{\sin 2\eta}{2\eta}+\frac{2}{d}S_2^2(\nu^2-\mu^2)\frac{\sin 2\mu}{2\mu }\nonumber\\
&+&\frac{d}{2\nu}S_3^2\sinh 2\nu+\frac{8\nu S_1S_2}{d(\eta^2-\mu^2)}\left [(\eta^2\Gamma\p-\mu^2\Gamma)\sin \eta \cos\mu-\eta\nu (\Gamma\p-\Gamma)\sin\mu\cos\eta \right ]\nonumber\\
&-&\frac{4S_1S_3}{\eta^2+\nu^2}\left [
(\nu^2\Gamma+\eta^2\Gamma\p)\cosh\nu\sin\eta+\nu\eta
(\Gamma\p-\Gamma)\sinh\nu\cos\eta\right ]-4S_2S_3\sinh\nu\cos\mu
\;\;.\label{A1}
\end{eqnarray}

For the EPS we are led to:

\begin{eqnarray}
M^{(EPS)}_{\kappa\,,\,n}&=&\frac{2}{d}S_1^2(\Gamma^2\nu^2+\Gamma^{'2}\eta^2)+\frac{2}{d}S_2^2(\mu^2+\nu^2)-\frac{2}{d}S_1^2(\Gamma^{'2}\eta^2-\Gamma^2\nu^2)\frac{\sin 2\eta}{2\eta}-\frac{2}{d}S_2^2(\nu^2-\mu^2)\frac{\sin 2\mu}{2\mu }\nonumber\\
&+&\frac{d}{2\nu}S_3^2\sinh 2\nu-\frac{8\nu S_1S_2}{d(\eta^2-\mu^2)}\left [(\eta^2\Gamma\p-\mu^2\Gamma)\sin \mu \cos\eta-\eta\nu (\Gamma\p-\Gamma)\sin\eta\cos\mu \right ]\nonumber\\
&+&\frac{4S_1S_3}{\eta^2+\nu^2}\left [
(\nu^2\Gamma+\eta^2\Gamma\p)\sinh\nu\cos\eta-\nu\eta
(\Gamma\p-\Gamma)\cosh\nu\sin\eta\right ]-4S_2S_3\cosh\nu\sin\mu
\;\;.\label{A2}
\end{eqnarray}

The other quantity is $F(z)$, which can be casted as:

\be \label{A3} F^{(OPS)}_{\kappa\,,\,n}(z)=\left
\{\begin{array}{lll}
S_1\sin(Q_Lz)+\frac{\beta_T^2(Q_L^2+\kappa^2)}{\kappa (\omega_L^2-\omega_T^2)}S_3\sinh(\kappa z)& \mbox{for}& -d/2<x<d/2\\
\exp(\nu)\left [
S_1\sin(\eta)+\frac{\beta_T^2(Q_L^2+\kappa^2)}{\kappa
(\omega_L^2-\omega_T^2)}S_3\sinh(\nu)  \right
]\frac{z}{|z|}\exp(-\kappa |z|)&\mbox{for}& |z|>d/2 \end{array}
\right. \ee for the OPS, and: \be \label{A4}
F^{(EPS)}_{\kappa\,,\,n}(z)=\left \{ \begin{array}{lll}
S_1\p\cos(Q_Lz)-\frac{\beta_T^2(Q_L^2+\kappa^2)}{\kappa (\omega_L^2-\omega_T^2)}S_3\p\cosh(\kappa z)&\mbox{for}& -d/2<z<d/2\\
\exp(\nu)\left[S_1\p\cos(\eta)-\frac{\beta_T^2(Q_L^2+\kappa^2)}{\kappa
(\omega_L^2-\omega_T^2)}S_3\p\cosh(\nu)   \right]\exp(-\kappa
|z|)& \mbox{for} & |z|>d/2
\end{array}\right.
\ee for the EPS.

\end{document}